\documentclass[
 reprint,
superscriptaddress,
 amsmath,amssymb,
 aps,
prb,
]{revtex4-2}

\usepackage{graphicx}
\usepackage{dcolumn}
\usepackage{bm}
\usepackage{subfigure,multirow}
\usepackage{url,hyperref}
\usepackage[T1]{fontenc}
\usepackage{mathptmx}
\usepackage[marginal]{footmisc}
\graphicspath{ {./figures/} }

\begin{document}

\preprint{APS/123-QED}

\title{Two mass-imbalanced atoms in a hard-wall trap: Deep learning integrability of many-body systems}

\author{Liheng Lang}
\thanks{These two authors contributed equally to this work.}
\affiliation{%
Department of Physics and Key Laboratory of Optical Field Manipulation of Zhejiang Province, Zhejiang Sci-Tech University, Hangzhou 310018, China
}%
\author{Qichen Lu}
\thanks{These two authors contributed equally to this work.}
\affiliation{%
Department of Physics and Key Laboratory of Optical Field Manipulation of Zhejiang Province, Zhejiang Sci-Tech University, Hangzhou 310018, China
}%

\author{C. M. Dai}
\affiliation{%
Department of Physics and Key Laboratory of Optical Field Manipulation of Zhejiang Province, Zhejiang Sci-Tech University, Hangzhou 310018, China
}%
\author{Xingbo Wei}
\affiliation{%
Department of Physics and Key Laboratory of Optical Field Manipulation of Zhejiang Province, Zhejiang Sci-Tech University, Hangzhou 310018, China
}%
\author{Yanxia Liu}
\affiliation{%
School of Physics and Astronomy, Yunnan University, Kunming 650091, China
}%
\author{Yunbo Zhang}%
\email{ybzhang@zstu.edu.cn}
\affiliation{%
Department of Physics and Key Laboratory of Optical Field Manipulation of Zhejiang Province, Zhejiang Sci-Tech University, Hangzhou 310018, China
}%


\date{\today}

\begin{abstract}
The study of integrable systems has led to significant advancements in our understanding of many-body physics. We design a series of numerical experiments to analyze the integrability of a mass-imbalanced two-body system through energy level statistics and deep learning of wavefunctions. The level spacing distributions are fitted by a Brody distribution and the fitting parameter $\omega$ is found to separate the integrable and non-integrable mass ratios by a critical line $\omega=0$. The convolutional neural network built from the probability density images could identify the transition points between integrable and non-integrable systems with high accuracy, yet in a much shorter computation time. A brilliant example of the network's ability is to identify a new integrable mass ratio $1/3$ by learning from the known integrable case of equal mass, with a remarkable network confidence of $98.78\%$. The robustness of our neural networks is further enhanced by adversarial learning, where samples are generated by standard and quantum perturbations mixed in the probability density images and the wavefunctions, respectively.

\end{abstract}
\maketitle

\section{Introduction}
Machine learning has been widely applied to big data, digital image processing, medical and other fields recently \cite{LeCun2015}, which aims at analyzing and processing new data through training machine with a great deal of data. Therefore, a series of ideas emerged in the field of physics, that is, whether physical data or formulas can be used as the input of machine learning to study the physics (such as quantum many-body and quantum topology). These ideas provide new approaches for the research of quantum physics. At present, the application of machine learning methods in physics \cite{Carleo2019} mainly includes classifying material phases and finding phase transition points \cite{Subir2011,Juan2017,Deng2017}, reconstructing wave functions \cite{Carrasquilla2021}, and solving physical equations \cite{Wei2018}, etc.

It was first proposed to solve the dimensionality curse in numerical calculation by deep learning method \cite{Goodfellow2016} to study the phase transition in a variety of condensed matter Hamiltonians \cite{Juan2017}. Two modern deep learning architectures, including full connection neural network and convolutional neural network, are built to analyze the phase transition between the ferromagnetic phase and antiferromagnetic phase of the Ising model from two complementary aspects. Deep learning was then applied to XY model \cite{Zhang2019,Stroev2021,Song2022}, Heisenberg model \cite{L2017} and other physical systems \cite{Dong2019,Venderley2018,Bartok2018}, which proves that neural networks have achieved some satisfactory results in solving complex physical systems, and also suggests that the combination of deep learning and physics is promising. In the study of the phase transition, unlike the traditional method of solving the order parameter in condensed matter physics, deep learning extracts feature information from a large number of physical data. In addition, some unsupervised methods have made major breakthroughs in phase transitions, such as principal component analysis (PCA) \cite {Wang2016,Wang2017,Wang2018}, restricted boltzmann machines (RBM) \cite{Torlai2016,Barra2017}, autoencoder \cite{Wetzel2017}, etc. In further analysis of the robustness of neural networks \cite{Lu2020,Jiang2023}, it was found that both classical classifiers and quantum classifiers are vulnerable to adversarial disturbances, which provokes new discussions in the interdisciplinary study of deep learning and physics. Neural network has very successful applications in the classification of matter phases, and also greatly promotes the research of quantum chaos \cite{Kharkov2020}. For example, it can learn the transition from chaos to integrable states by analyzing the wavefunction and reveal the features of quantum chaos \cite{Huber2021}.

Most of the exactly solvable models are limited to the case of equal mass for one-dimensional quantum many-body systems. However, researches show that there exist several types of few-body systems with imbalanced masses \cite{Olshanii2015,Dehkharghani2016,Harshman2017,Olshanii2016,Liu2019,Liu2021,Huber2021} or with a $\delta$-barrier in the potential center \cite{Liu2015,Gomez2019} that are found to be solvable for some specific mass ratios or in the spatially odd sector of the Hilbert space. We focus on the two-particle system with arbitrary interaction strength and mass imbalance in an one dimension hard wall trap. In previous work, we accurately solved the Bethe-type ansatz equation and found that the system is integrable \cite{Liu2019} when the mass ratios $\eta$ is $1$ or $1/3$. Though there are various definitions of quantum integrability, we more specifically adopt the Bethe-ansatz integrablity, i.e. any eigenstate can be written as a finite superposition of plane waves, while the non-integrable states are often called chaotic states. In this paper, we design a series of numerical experiments to analyze the integrability of mass imbalanced many-body systems through statistics of the energy spectrum and deep learning of wavefunctions. For the energy level statistics, we take the non-interacting wavefunctions as basis vectors to calculate the system energy by the method of exact diagonalization (ED), determine the statistical properties of the unfolded spectrum by means of the nearest neighbor level spacing distribution, and analyze the integrable and chaotic states of the quantum system for different mass ratios. In the deep learning stage, we build a convolutional neural network that can implement binary classifications, and include into the dataset the wavefunctions of high excited states for both integrable and non-integrable systems. The neural network is trained by a few iterations and the accuracy achieves 99$\%$. The trained model is then implemented to predict the system integrablility for other mass ratios.

Our objectives are twofold: firstly, to accurately classify input states into either integrable or non-integrable, and secondly, to test the robustness of the neural network under perturbations. We then investigate the impact of noise on neural networks by generating adversarial samples by two methods, the standard perturbation in the density probability images, and quantum perturbation in the wavefunctions. Standard perturbation allows the neural networks to extract features from integrable and non-integrable density images, while quantum perturbation may help in exploring unique states intrinsic in physical systems. Adversarial samples are generated from both perturbations to test the vulnerability of the original network. We aim to enhance the network's robustness through adversarial training and improve the accuracy of the original network by incorporating quantum adversarial samples into the learning process.

The rest of the paper is organized as follows. In Section II, we describe the two-particle interacting 1D model with imbalanced mass and numerically generate the energy spectrum of the system for different interaction strengths and mass ratios. To secure the ED method, we estimate the accuracy of numerical results for the energy levels by directly comparing the exact results with  the Bethe-ansatz equations. We present our results of energy level statistics, deep learning of wavefunctions, and adversarial learning in Section III. Detailed procedure is given for dataset production, neural network construction, and numerical experiments. The standard and quantum perturbations are respectively carried out for the adversarial training. Section IV summarizes our findings. 

\section{Model}

We consider an interacting system of two particles with masses $m_1$ and $m_2$ confined in a 1D hard-wall trap of length $L$. The atoms interact with each other through contact $\delta$-interaction and the Hamiltonian of system can be written as
\begin{equation}
  H=-\frac{\hbar^2}{2 m_1} \frac{\partial^2}{\partial x_1^2}-\frac{\hbar^2}{2 m_2} \frac{\partial^2}{\partial x_2^2}+g \delta\left(x_1-x_2\right),
  \label{Hamiltonian1}
\end{equation}
where $x_1$ and $x_2$ are coordinates of the two particles and $g$ is the interaction strength. Denoting the mass ratio of the two particles as $\eta =m_1 / m_2$ (note that $\eta$ defines the same Hamiltonian as $1/\eta$ does thus we set $\eta<1$). Classically each collision and reflection process of the two particles gives rise to a new pair of momenta. Though a closed set of finite numbers of the momentum may exist for $\eta$ satisfying the nonergodicity condition $\eta=\tan ^2 l \pi / 2 n$ with $l$ and $n$ positive integers, it is found in previous work that the system is integrable only when the mass ratio is $1$ and $1/3$ \cite{Liu2019}. In the case of equal mass for $\eta = \tan^2 \pi/4 = 1$, the full momentum set consists of 8 elements which form a dihedral group $D_4$. When $\eta  = \tan^2 \pi/6 = 1/3$, the collision operators form a dihedral group $D_6$, which is the only exactly solvable example of quantum mass-imbalance systems. We exactly solved this interacting two-particle system in a hard-wall with mass ratio $\eta=1/3$ with the Bethe Ansatz hypothesis for the wave function, which is generalized to the so-called asymmetric Bethe Ansatz (ABA) method in Ref.\cite{Jackson2023} and our case is proven to be a particular instance. For $\eta$ other than $1$ and $1/3$, we may still find the corresponding dihedral groups $D_{2n}$, however, the quantum system is not integrable and the trajectory of two particles is chaotic and unpredictable. Based on these previous works, this paper puts forward studying the influence of mass ratio of the two particles on system integrability with the purpose of exploring transitions between integrable and chaotic. 

To numerically generate the system energy levels, we take the product of two single-particle wave functions in a hard-wall trap of unit length $L=1$ as the basis vectors, which can be written as
\begin{equation}
  \psi_{ij}\left(x_1, x_2\right)=2 \sin i \pi x_1 \sin j \pi x_2,
  \label{basis}
\end{equation}
where $i$ and $j$ are positive integers, and $x_1$ and $x_2$ denote the coordinates of the lighter and heavier particles, respectively. The matrix elements of the Hamiltonian (\ref{Hamiltonian1}) on the basis (\ref{basis}) are written as (set $\hbar=1$)
\begin{equation}\label{matrixelement}
\begin{aligned}
  &H_{i_1j_1,i_2j_2}=4 \int_0^1 \int_0^1 d x_1 d x_2 \sin i_1 \pi x_1 \sin j_1 \pi x_2 \\
  &\left[-\frac{1}{2m_1} \frac{\partial^2}{\partial x_1^2}-\frac{1}{2m_2} \frac{\partial^2}{\partial x_2^2}+g \delta\left(x_1-x_2\right) \right]\sin i_2 \pi x_1 \sin j_2 \pi x_2.
\end{aligned}
\end{equation}
A straightforward calculation gives the matrix elements of the Hamiltonian
\begin{equation}
  H_{i_1j_1,i_2j_2}=\pi^2 \frac{i_1^2 m_2+j_1^2 m_1}{2m_1 m_2} \delta_{i_1,i_2}
  \delta_{j_1,j_2}+\frac{g}{2}I,
  \label{HamiltonianD}
\end{equation}
where $\delta_{i,j}$ is the discrete $\delta$-function, and $I$ is defined as
\begin{equation}
\begin{aligned}
  I&=\delta_{i_1-i_2+j_1-j_2,0}
  +\delta_{i_1-i_2-j_1+j_2,0}\\
  &-\delta_{i_1-i_2+j_1+j_2,0}
  -\delta_{i_1-i_2-j_1-j_2,0}\\
  &-\delta_{i_1+i_2+j_1-j_2,0}
  -\delta_{i_1+i_2-j_1+j_2,0}\\
  &+\delta_{i_1+i_2+j_1+j_2,0}
  +\delta_{i_1+i_2-j_1-j_2,0}.
\end{aligned}
\end{equation}
The Hamiltonian is then solved numerically for arbitrary mass ratio $\eta$ and interaction strength $g$ by the method of ED. In the numerical procedure, the mass ratio is taken as the only variable such that the two particle masses are related to $\eta$ as
\begin{equation}
    m_1=1+\eta, \qquad m_2=1+1/\eta,
    \label{mass}
\end{equation}
which leads to a unity reduced mass $\mu=1/(\frac{1}{m1}+\frac{1}{m2})=1$. The energy level spectrum is analyzed statistically to find the distribution of the level spacing and the wave functions serve as the training set for our neural network. Notice that the non-zero off-diagonal matrix elements is determined by the relations between the four integers $(i_1,i_2,j_1,j_2)$. Together with the diagonal elements, we find the number of non-zero matrix elements grows as $\sim N^3$ for large enough $N$, where $N$ is the cut-off number of the non-interacting basis vectors. Since we use the eigenstates of the non-interacting system as the basis vectors, one may expect ED to perform best for small values of $g$ and more poorly for large $g$. 

To estimate the accuracy of our results, we benchmark against the exact solution for $\eta=1/3$ which is exactly solvable. Using $N = 230$, we compute the 5000 lowest levels of the system. This set will be used as input for our analysis in the next section.

\begin{figure}[]
    \includegraphics[width=8.6cm]{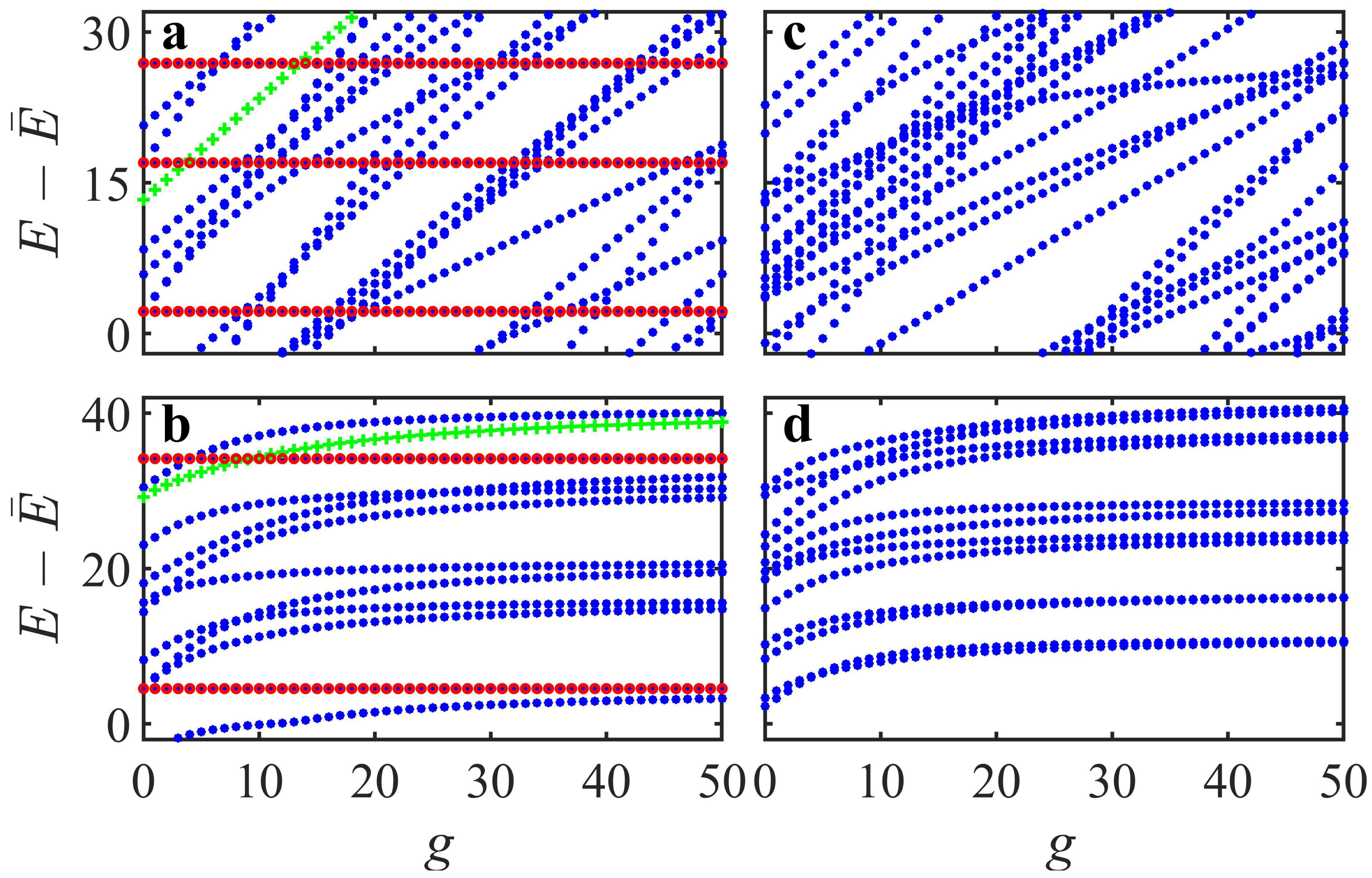}
    \caption{{The energy spectrum of integrable and nonintegrable systems. The left panels show the higher (a) and lower (b) energy spectrum for an integrable system $\eta=1/3$ with horizontal energy levels (red). The right panels show the higher (c) and lower (d) spectrum for non-integrable system $\eta=3-2\sqrt{2}$. The first two horizontal levels, with level numbers 8 and 18 for $g=0$, are shown in (b), while three with higher level numbers 4818, 4826 and 4830 are shown in (a) indicated by red dots. All energy levels are plotted upon a shift energy $\bar E$ which are 13341.5(a), 10909(c), 30(b) and 20(d), respectively. Levels noumbers 16 and 4825, denoted in green pluses, are chosen to show the accuracy of the numerical results compared with the Bethe Ansatz exact results in Fig. \ref{error}. }}\label{spectrum}
\end{figure}

In previous work \cite{Liu2019}, it is found that for $\eta=1/3$ the energies of some special states, e.g., the 8th level, do not change with the interaction strength $g$ and the existence of such states is attributed to the emergence of a triple degenerate point in the non-interacting limit $g = 0$. From the matrix elements in the non-interacting basis, it is easy to understand what happens in the degenerate point as follows: in some cases three pairs of integers $(i,j)$ give the same value of the expression $3i^2+j^2$, in which case one can always find an eigenvalue independent of $g$, although not all $\delta$ functions in $I$ vanish. These interaction-independent levels are exactly constants and serve as perfect benchmarks for estimating numerical accuracy. We find that there are 649 horizontal levels in the first 5000 levels. The first one appears as the 8th level and does not cross with any other, but the second horizontal level, whose level number is 18 at $g=0$, has already been crossed twice as the interaction becomes strong enough, ranking the 16th level at $g=50$, as shown in Fig. \ref{spectrum}b. This level crossing makes it difficult to identify the ordering of the levels, especially for higher energy and strongly interacting case. In Fig. \ref{spectrum}a we show the spectrum between three adjacent horizontal levels, namely level numbers 4816, 4826 and 4830, ordered in the non-interacting case. Clearly these high levels increase more rapidly with the interaction than lower levels, and level crossing occurs more often for higher energy levels and larger $g$. Note that the horizontal levels disappear immediately for other mass ratios, e.g. $\eta=3-2\sqrt{2}$ in Fig. \ref{spectrum}d for lower levels and Fig. \ref{spectrum}b for higher levels due to the non-integrability of the system. Note that we have deliberately show the higher and lower levels in a comparable range such that all levels are plotted upon an energy shift $\bar{E}$ which are different for higher or lower levels.

\begin{figure}[]
\includegraphics[width=8.6cm]{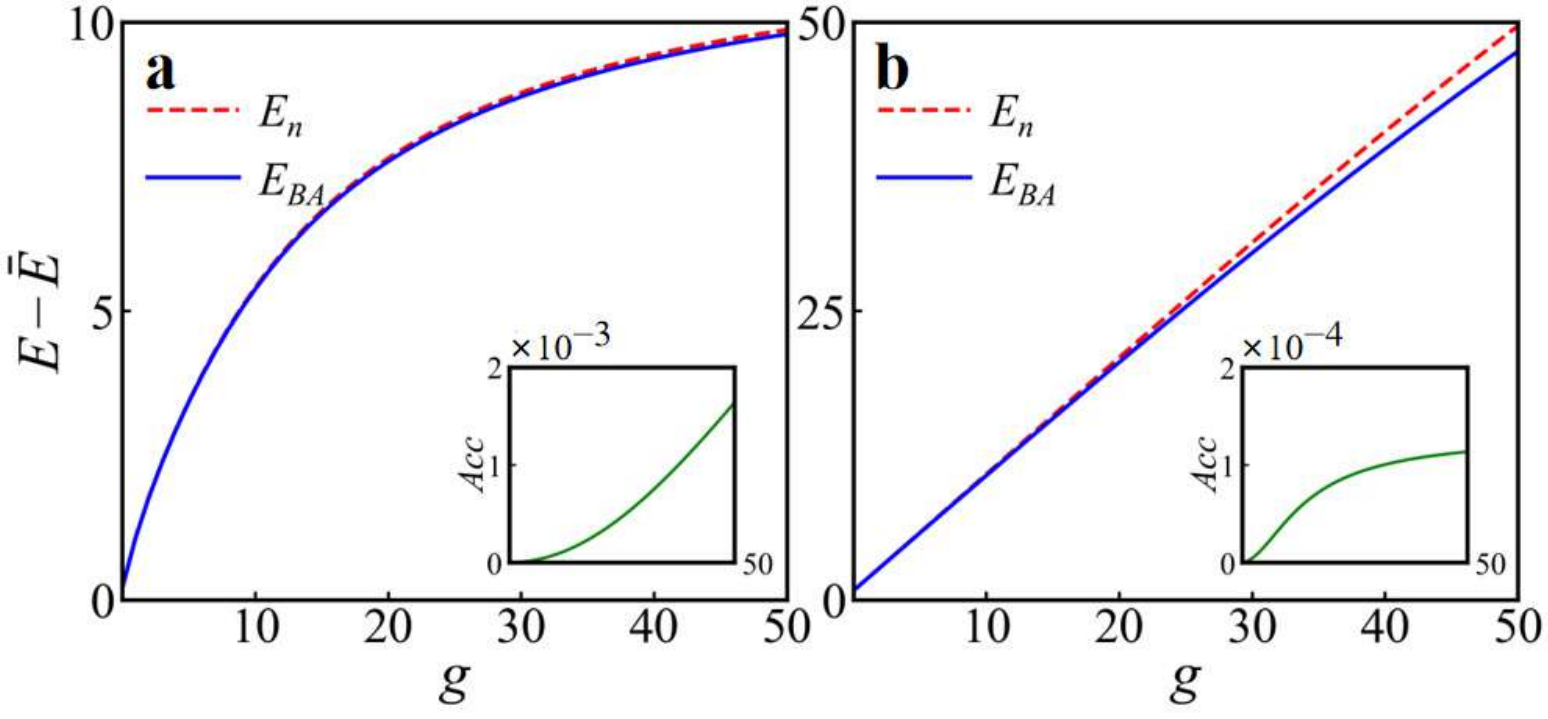}
    \caption{{The lower (level number 16, left) and higher (level number 4825, right) energy levels calculated by numerical method (red dashed lines) and BA exact method (blue solid lines). Insets: the relative accuracy $Acc$ as a function of the interaction strength $g$ for the lower and higher levels. The energy shifts $\bar{E}$ of (a) and (b) are 59 and 13354 respectively.}}\label{error}
\end{figure}

We estimate the accuracy of numerical results by directly comparing the numerical values for the energy levels $E_n$ at finite $g$ with the exact results $E_{BA}$ from the BA equations in the $\eta=1/3$ case. We search the exact BA results from the requirement that for any interaction strength $g$ the momentum $k_1, k_2$ should satisfy the Bethe Ansatz equations (BAEs)
\begin{eqnarray}
k_2+3k_1-2g\left(\cot(k_2+k_1)/2 +\cot k_1\right)=0,\label{BAE1}\\
k_2-3k_1-2g\left(\cot(k_2-k_1)/2 -\cot k_1\right)=0,
\label{BAE2}
\end{eqnarray}
for even parity or
\begin{eqnarray}
k_2+3k_1+2g\left(\tan(k_2+k_1)/2 +\tan k_2\right)=0,\label{BAE3}\\
k_2-3k_1+2g\left(\tan(k_2-k_1)/2 -\tan k_1\right)=0,
\label{BAE4}
\end{eqnarray}
for odd parity in the natural units $\hbar = \mu = L = 1$. It is convenient to start the search from the non-interacting side $g=0$. We first fix the two integers $(i,j)$ at $g=0$, e.g. by sorting the levels for increasing $i$ and $j$ one by one. For lower level, e.g. level number 16, it is easy to see $(i,j)=(2,6)$, while for higher level, e.g. level number 4825, we have $(i,j)=(60,5)$. Note that the horizontal levels $E_h$ are triple degenerate, which for instance gives us three pairs of $(i,j)=(53,49), (51,55), (2,104)$ for level numbers 4826, 4827, 4828. The interaction alters the values of $(i,j)$ into non-integers with the energy given by $E_{BA}=(3k_1^2+ k_2^2)/8$. For each finite value $g$, we numerically find the solution for $k_1$ and $k_2$ by solving the BAEs (\ref{BAE1},\ref{BAE2}) or (\ref{BAE3},\ref{BAE4}) in the momentum range $(k_1,k_2)=(i\pi\pm \Delta k,j\pi\pm\Delta k)$. We take $\Delta k=1$ for all calculations as the momentum values for each level will not change drastically even for very strong interaction. The solutions need to be refined by the scattering relations 
\begin{equation}
(k'_1,k'_2)^T=-\sigma_z s(1/3) \sigma_z (k_1,k_2)^T 
,\end{equation}
and
\begin{equation}
(k''_1,k''_2)^T=\sigma_z s(1/3)(k_1,k_2)^T, 
\end{equation}
with
\begin{equation}
    s(1/3)=\left[ \begin{matrix}
-\frac{1}{2} & \frac{1}{2} \\
\frac{3}{2} & \frac{1}{2}
\end{matrix} \right],\qquad
\sigma_z=\left[ \begin{matrix}
1 & 0 \\
0 & -1
\end{matrix} \right].
\end{equation}

\begin{figure}[]
    \includegraphics[width=8.6cm]{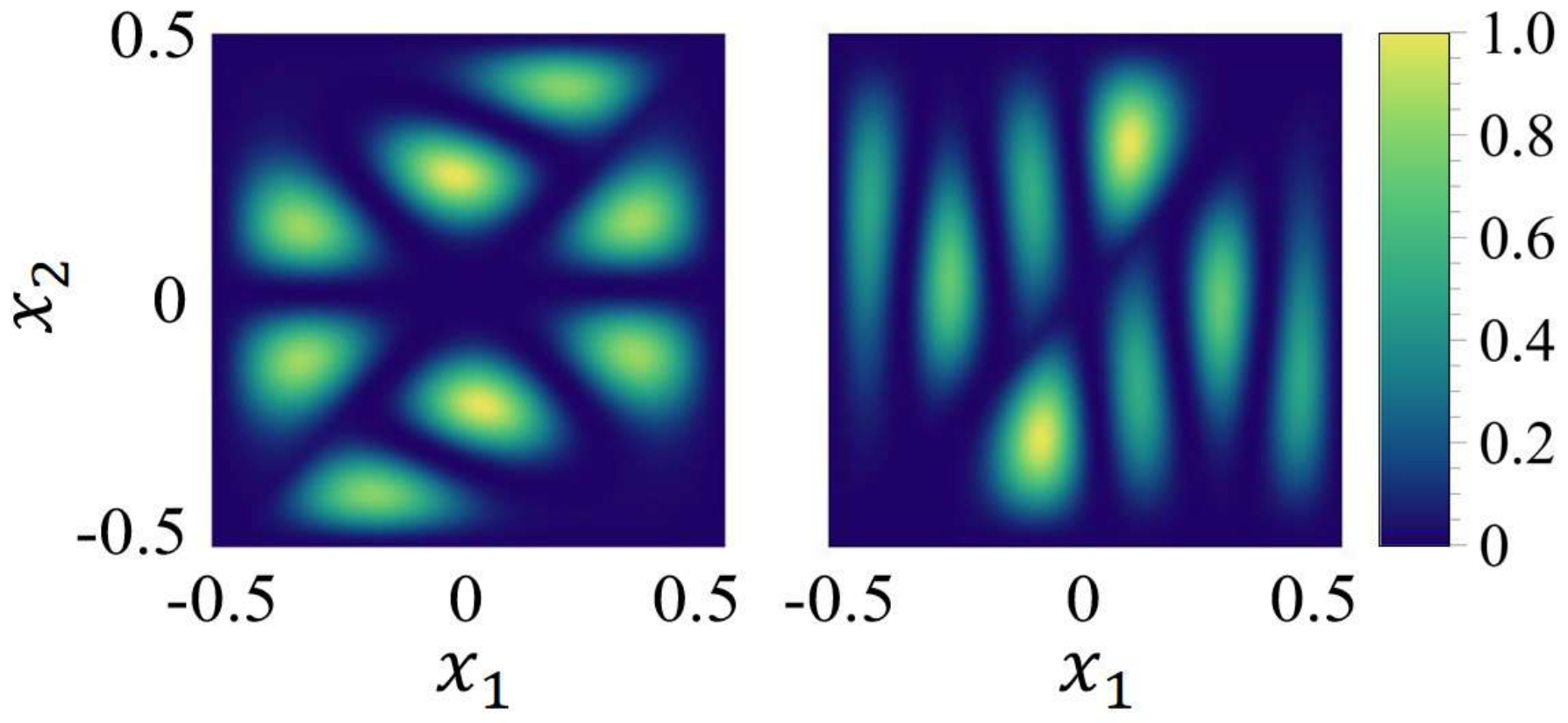}
    \caption{{The wave function of the $10$th state for an integrable system ($\eta=1/3$, left) and that for an non-integrable system ($\eta=3-2\sqrt{2}$, right). The eigen energies are $E_{10}=44.90$ and $E_{10}=35.39$ respectively. Note in both integrable and non-integrable cases, there exist many nodal lines in the density distributions. }}\label{wavefunction}
\end{figure}

It is necessary that the momentum values $(k'_1,k'_2)$ after one scattering and $(k''_1,k''_2)$ after two scattering processes still satisfy the BAEs. In Fig. \ref{error} we show the numerical and exact results and the difference between them for level numbers 16 and 4825, indicated in Fig. \ref{spectrum} by green crosses. Clearly the ED method works well for lower level and small values of $g$. The relative accuracy, defined as $Acc=|E_n-E_{BA}|/E_{BA}$, however, is still within the order of $10^{-3}$ for high level number around 5000 and large $g=50$.

\begin{figure}
  \centering
  \includegraphics[width=8.6cm]{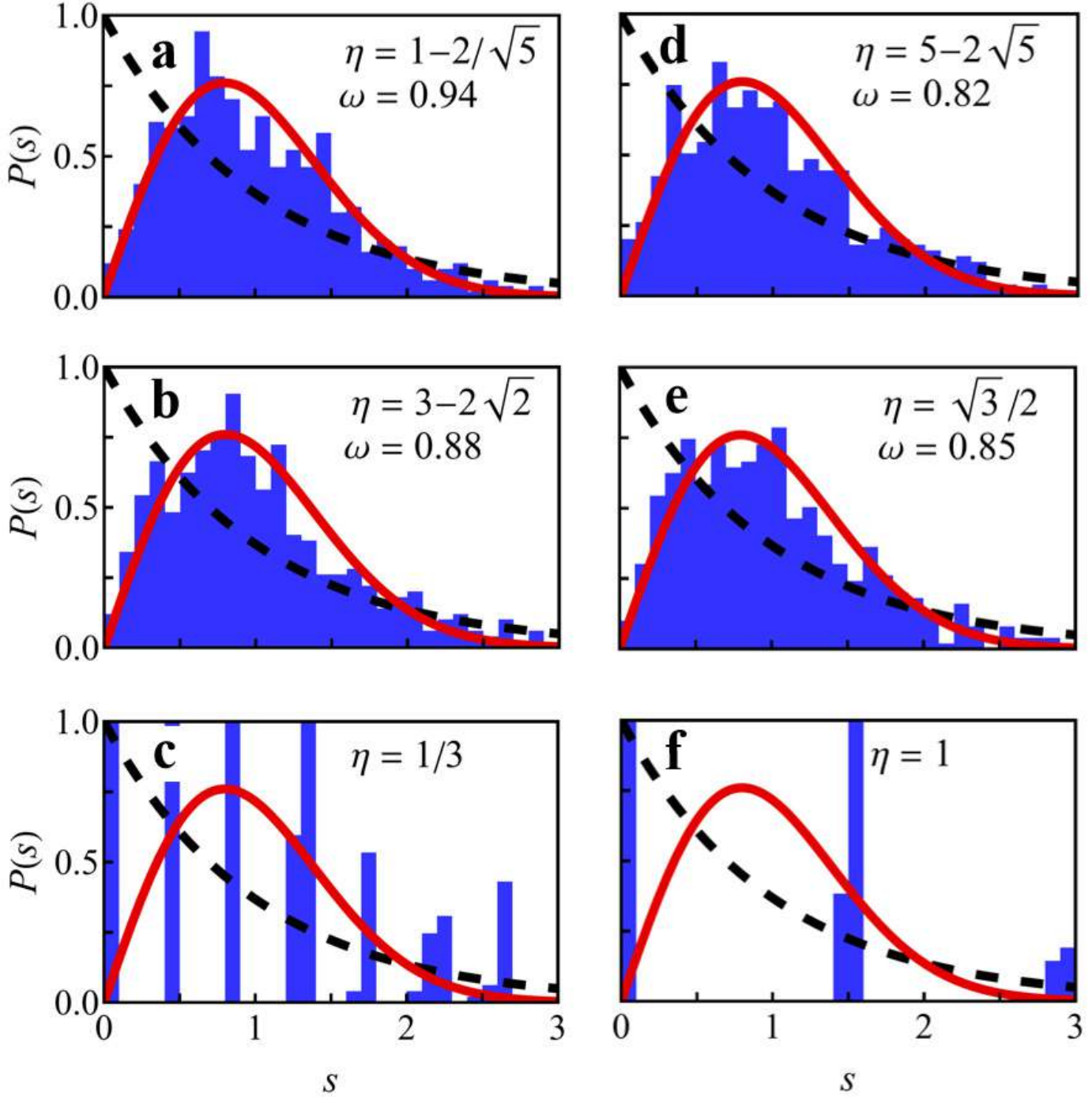}
  \caption{{The histogram of nearest neighbor energy level distribution $P(s)$ for different mass ratios $\eta$ of two particles at $g=0$. The red solid curve shows the Wigner distribution while the black dashed curve shows the Poisson distribution. The fit with a Brody distribution gives the quoted $\omega$ in each panel. To produce these figures, only states with odd parity are used. The panels $a,b,d,e$ are chaotic, and the panels $c,f$ are integrable. Due  to the large number of degenerate levels, integrable states show sparse well-spaced distribution, while non-integrable states are close to Wigner distribution.}}\label{histogram}
\end{figure}
  
It has been proposed that the wave functions of integrable states are expected to have some non-trivial morphology, since the classical phase space of integrable systems has some interesting structure \cite{Huber2021}. The probability density of the wave function reflects the probability of particles appearing at different positions in the phase space, which carries important information about the integrability of the system \cite{Harshman20171}. We show in Fig. \ref{wavefunction} the probability density of the $10$th state for an integrable ($\eta=1/3$) and that for an non-integrable ($\eta=3-2\sqrt{2}$) system. The decisive feature of these states, whether integrable or not, is the existence of many nodal lines in the density distribution, which are defined as sets of points ${x_1, x_2}$ that satisfy $\psi(x_1, x_2) = 0$. The numerical data of density distribution produced from ED method are so large that it is hopelessly difficult for our eyes to find universal patterns of integrable system. Hence these data will be used as the input in our neural network for a progressive learning of features from the input data, which will be discussed in Section 3.

\section{Methods and results}
\subsection{Energy level statistics}

To analyze the statistical fluctuations of the spectrum $\{E_n\}$, it is necessary to separate the fluctuating part from its smoothed average part, the behavior of which is non-universal and cannot be described by random-matrix theory \cite{Alhassid1990}. We thus construct the staircase function of the spectrum $N(E)$, defined as the number of levels below $E$ \cite{Dong2021,Karampagia2015,Alhassid1992}, and separate it into average part $N_{av}(E)$ and quantum fluctuation. The non-universal part $N_{av}(E)$ may be taken to be the fit of a smooth function to the staircase $N(E)$, while the statistical fluctuations of the spectrum are found to be quite independent of the fitting class. Here we shall use fits with a polynomial of sixth order \cite{Alhassid1992}. The unfolded spectrum is defined by the mapping $\tilde E_n=N_{av}(E_n)$, and the unfolded levels $\tilde E_n$ have a constant average spacing.

\begin{figure}
  \centering
  \includegraphics[width=8.6cm]{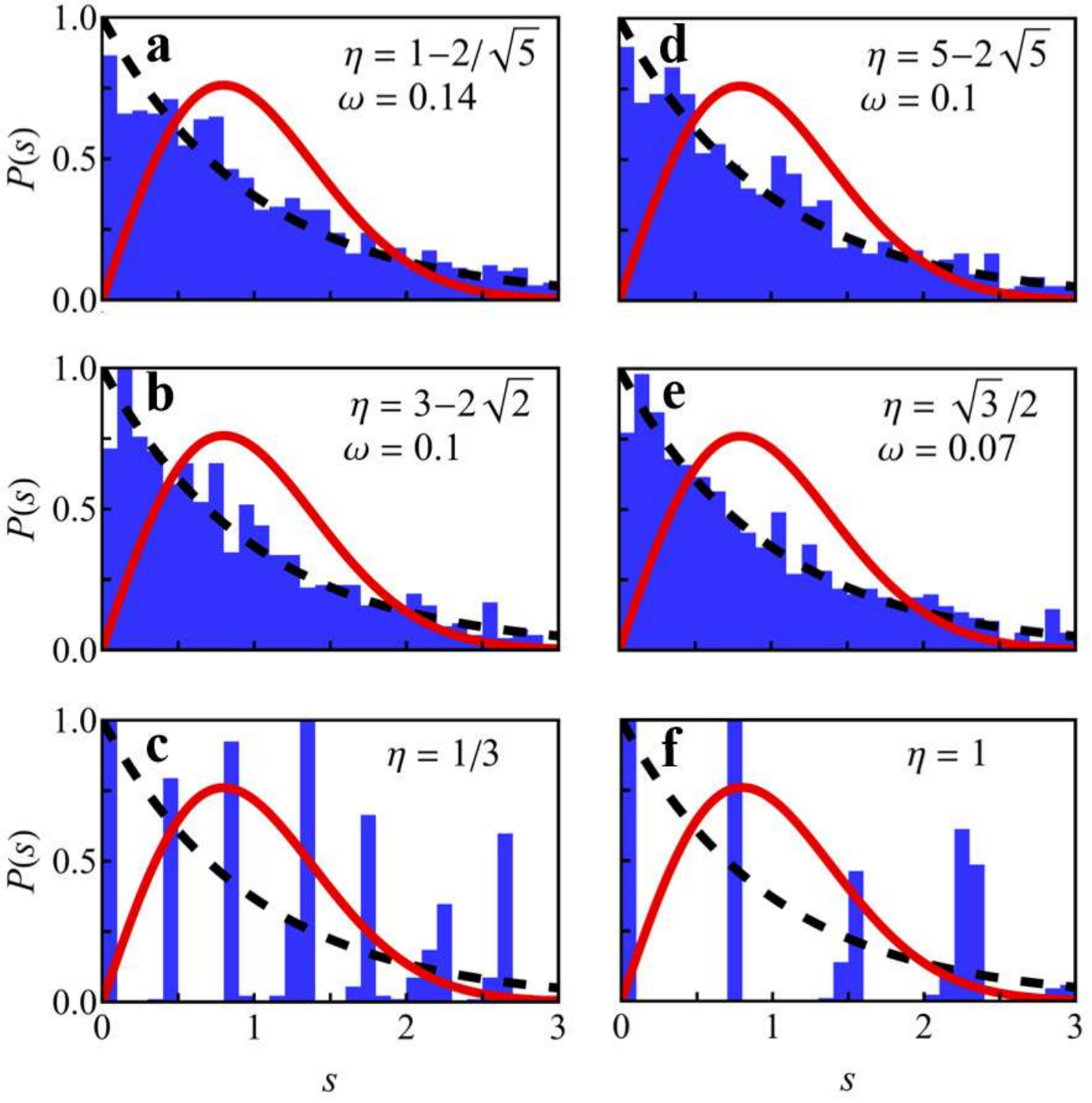}
  \caption{{The histogram of nearest neighbor energy level distribution $P(s)$ for different mass ratios $\eta$ of two particles at $g=0$. The red solid curve shows the Wigner distribution while the black dashed curve shows the Poisson distribution. The fit with a Brody distribution gives the quoted $\omega$ in each panel. This figure makes no distinction between parity. The panels $a,b,d,e$ are chaotic with much smaller values of $\omega$ than those in Fig. \ref{histogram}, and the panels $c,f$ are integrable. There is still a clear difference between integrable and non-integrable states.}}\label{histogram_1}
\end{figure}

\begin{figure}
  \centering
  \includegraphics[width=8.6cm]{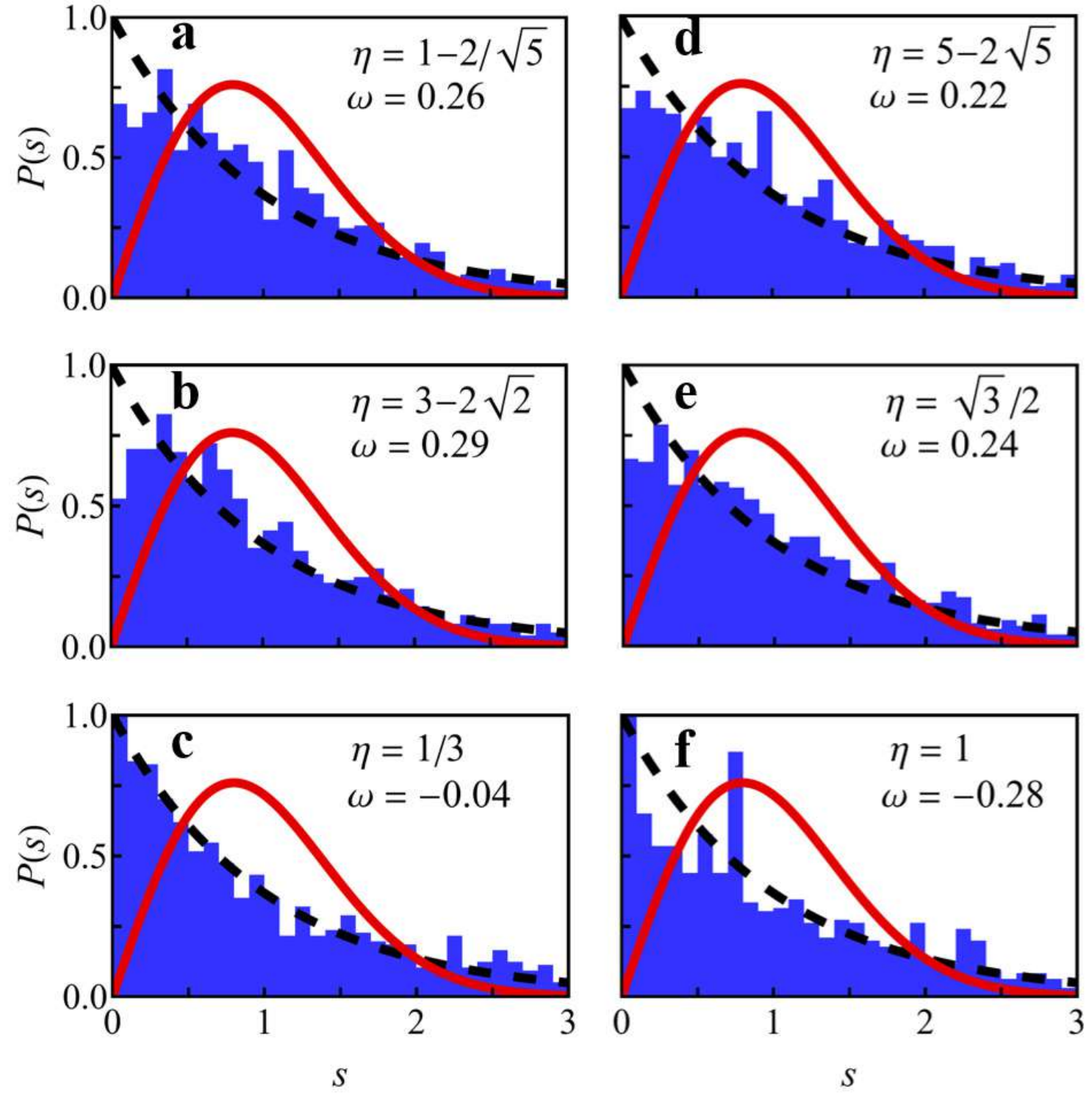}
  \caption{{The histogram of nearest neighbor energy level distribution $P(s)$ for different mass ratios $\eta$ of two particles at $g=20$. The red solid curve shows the Wigner distribution while the black dashed curve shows the Poisson distribution. This figure makes no distinction between parity. The panels $a,b,d,e$ are chaotic with positive $\omega$ ranging between $0.2$ and $0.3$, with the peaks appearing at non-zero values of  $s$, and the panels $c,f$ for integrable cases are Poisson with the maximum value occurring at $s = 0$ and negative values of $\omega$.}}\label{histogram_2}
\end{figure}

We determine the fluctuation properties of the unfolded levels by means of the nearest neighbor level spacing distribution $P(s)$, which is defined as the probability of two neighboring levels apart by a distance $s$. While a integrable system is expected to be dominated by the Poisson statistics $P(s)=e^{-s}$ with eigenvalues uncorrelated, we expect to obtain the Wigner distribution $P(s)=(\pi/2) s e^{-\pi s^2/4}$ for a chaotic system. We calculate the spacings $s_n$ from the unfolded spectrum $s_n=\tilde E_{n+1}-\tilde E_n$ and show the histograms in Figs. \ref{histogram},\ref{histogram_1} and \ref{histogram_2}, for several values of mass ratio $\eta$ in the absence or presence of interaction.

The level spacing distributions are fitted by a Brody distribution, which is of the form \cite{Brody1981,Bohigas1984,Alhassid1990,Miltenburg1994}
\begin{equation}
  P(s)=\alpha(1+\omega) s^\omega e^{-\alpha s^{1+\omega}}
,\end{equation}
where $\omega$ is the fitting parameter, $P(s)$ is the probability that $s_n$ falls within the interval ($s$, $s+ds$), and 
\begin{equation}
    \alpha=\Gamma[(2+\omega) /(1+\omega)]^{1+\omega}
,\end{equation} 
with $\Gamma$ is Gamma function. In these histogram figures, we depict the standard Wigner distribution ($\omega=1$) with red solid curves and the Poisson distribution ($\omega=0$) with black dashed curves. The fit with a Brody distribution gives the quoted $\omega$ in each panel.

We first consider the level spacing distribution in the case $g=0$ with only odd parity states are used, as the case with only even parity states leads to qualitatively identical results. The number of bins are taken at approximately a square root of the number of the considered levels (here we take 1000 levels into the statistics such that 30 bins are used in the histograms). With the mass ratio varied in the range $0< \eta\le 1$, two integrable values $1/3$ and $1$ are encountered, for which the degeneracies in the energy spectrum lead to well-spaced bins as shown in Fig. \ref{histogram}c and \ref{histogram}f. The degeneracies for the same parity in the integrable cases lead to well spaced bins in the histograms which is rather unique. This behavior is immediately broken for other mass ratios and the levels tend to repel each other, and the distribution $P(s)$ are well approximated by the Wigner distribution with quoted $\omega$ very close to 1 as shown in Fig. \ref{histogram}a,b,d, and e. We find that the space between the nearest bins in the case of equal masses is larger than the $\eta=1/3$ case, implying the fitting parameter $\omega$ further away from the Wigner distribution $\omega=1$. Note that it is essential that we do the statistics for the same parity. If we put all levels into the statistics without distinction of the parity, as in Fig. \ref{histogram_1}, level distributions with well-spaced bins in the integrable cases survive for different parities, however, the Wigner distribution for non-integrable mass ratio cannot be well established. We observe clear deviation of the distribution away from the standard Wigner distribution with quite small values of $\omega \sim 0.1$ in the chaotic cases. 

When the interaction is turned on, for instance $g=20$, it is no longer possible to distinguish the parity of each level from our numerical ED results. While the level degeneracies in the spectrum are preserved for integrable cases due to the horizontal levels, they disappear for the aspect ratios other than 1 and $1/3$ (see Fig. \ref{spectrum}c and d). The histograms for the integrable cases conform to the Poisson distribution with the maximum value occurring at $s=0$ and negative values of $\omega$, and those for the non-integrable cases are Brody distributions with positive $\omega$ ranging between $0.2$ and $0.3$, with the peaks appearing at non-zero values of $s$, as shown in Fig. \ref{histogram_2}. The mass ratios in these histograms are in ascending order as $0.11, 0.17, 0.33, 0.53, 0.87, 1.00$ in figure panels (a) to (f), respectively. The best fit curve depends on the number of bins we choose, and we find that even for the same mass ratio, different bins result in different fit values $\omega$. The degeneracy of the energy levels leads to negative values of $\omega$, which is dubbed "overintegrable" because of large number of exact degeneracies \cite{Alhassid19921} and happens rather frequently in the study of chaos in the low-lying collective states of nuclei \cite{Whelan1993,Canetta2000,Hou2009,Wu2022}. This deviation from the random matrix theory due to the level degeneracies is our motivation for the study of system integrability with deep learning.

\begin{figure}
  \centering
  \includegraphics[width=8.6cm]{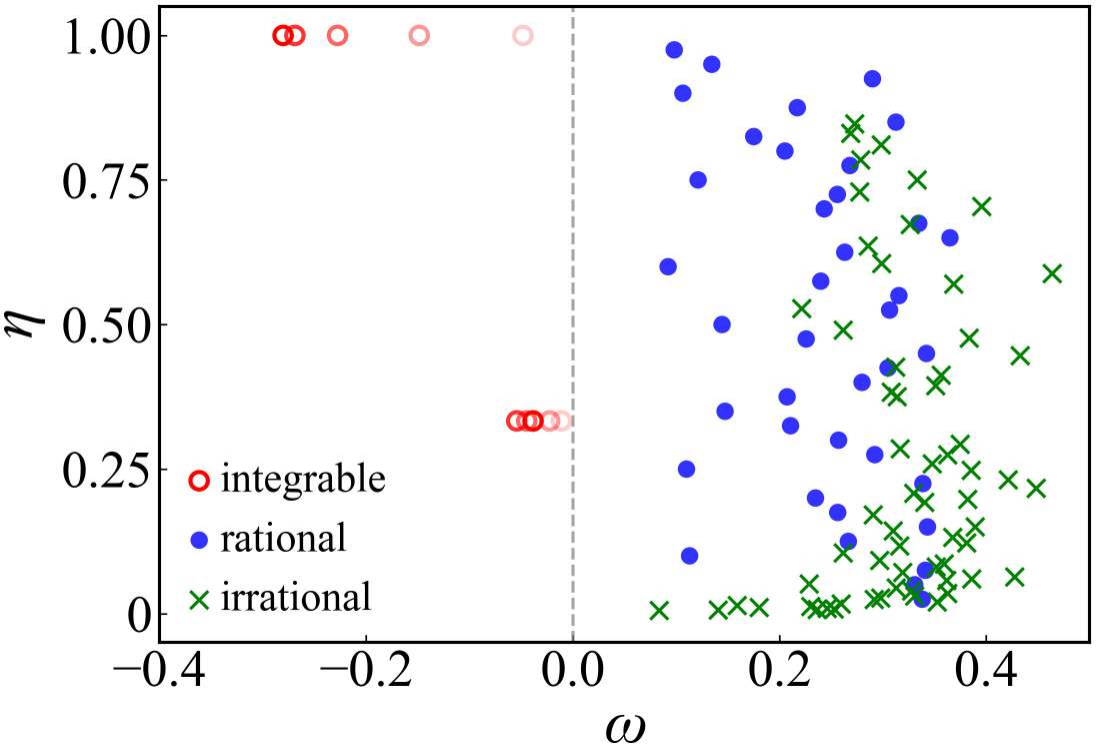}
  \caption{{The fitting parameters $\omega $ for different mass ratios. Red hollow points represent the integrable mass ratios, blue solid dots represent the rational non-integrable mass ratios and green cross are non-ergodicity points satisfying $\eta=\tan^2l\pi/2n$. The bins in the statistics for the integrable points are 10, 15, 20, 25, 30 (from light to deep), respectively. Clearly the boundary between integrable and non-integrable points is $\omega=0$.}}\label{massratio}
\end{figure}

Among the candidates for the mass ratio $\eta$ that fulfill the closeness of scattered momentum vector and the corresponding dihedral group of the collision operators, it was found that $\eta=1/3$ is the only exactly solvable example of quantum mass-imbalance systems by means of Bethe-type-ansatz method. To exclude other possible integrable values of mass ratios in the dihedral group, we do the level statistics for all values of $\eta$ satisfying the non-ergodicity condition (which are irrational for $4 \le n\le 20$) by taking the first $1000$ levels into statistics. In addition, we successively scan the values of the mass ratio by a step $0.025$ in the whole range $0<\eta<1$ (thus rational) and extract the fitting parameters sequentially. The resultant fitting parameters $\omega$ are shown in Fig. \ref{massratio} and we find there is an obvious critical line between integrable and non-integrable points $\omega=0$, i.e. all integrable points lie on the left with negative value of $\omega$, while all non-integrable points on the right with positive $\omega$. The bins in the statistics for the integrable points are varied between 10 and 30 for visual convenience and the values of $\omega$ remain negative. No new integrable points are found.

\begin{figure}
  \centering\includegraphics[width=8.6cm]{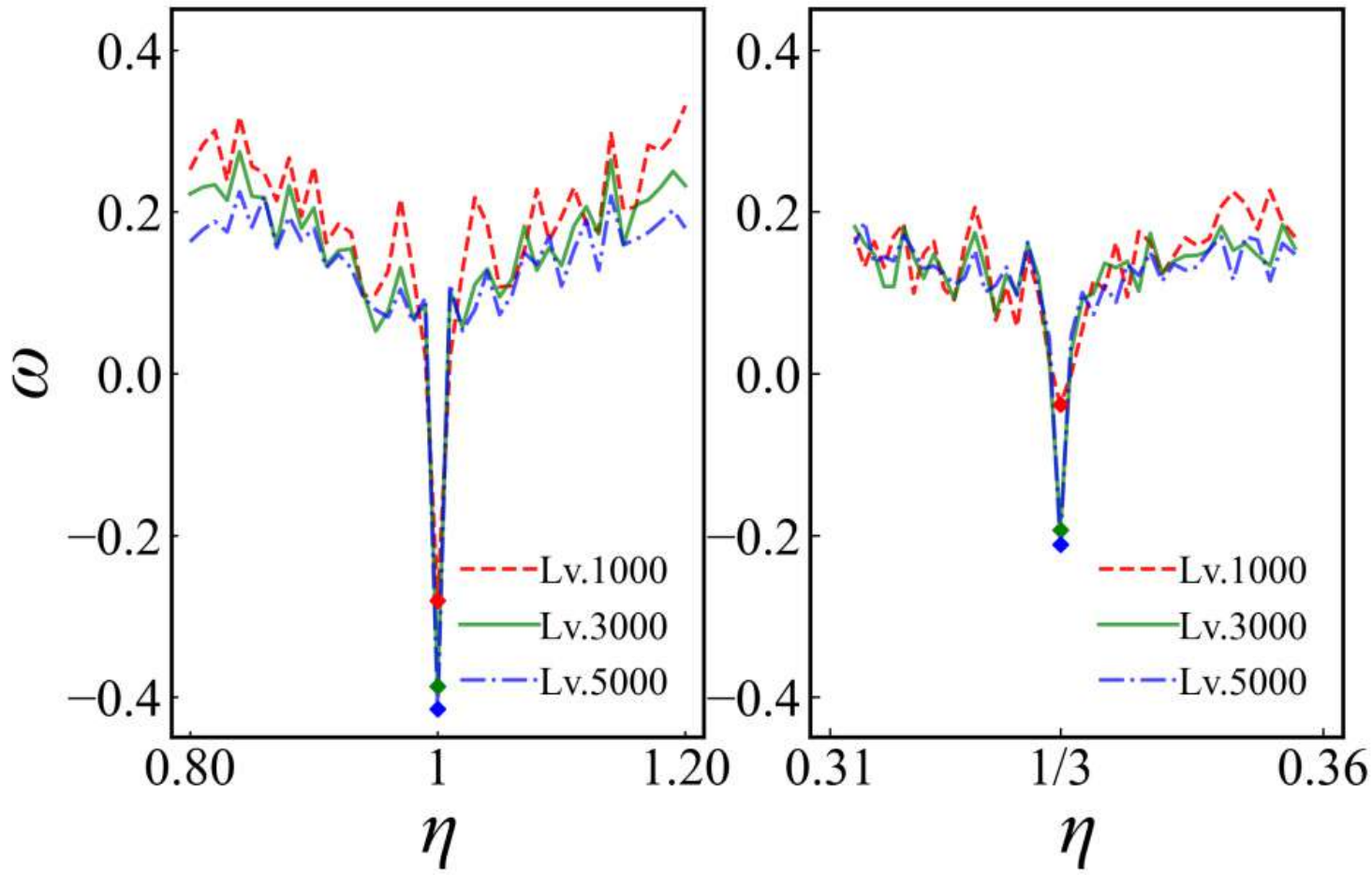}
  \caption{{The abrupt change of $\omega $ near the integrable points. (a) and (b) show the cases of $\eta=1$ and $\eta=1/3$ respectively. Red, green, and blue curves represent the same distribution fitting scheme for 2000, 3000, and 5000 energy levels, with the bins taken approximately as the square root of level numbers as 30, 50, and 70, respectively.}}\label{massratios}
\end{figure}

We further calculate the fitting parameter $\omega$ near the integrable points of the mass ratio by including more energy levels in the statistics, see Fig. \ref{massratios} for the two integrable points with the number of levels 1000, 3000, and 5000, with the bins taken approximately as
the square root of level numbers as 30, 50, and 70, respectively. It can be clearly seen that when the mass ratio deviates slightly from the integrable point, the values of $\omega$ immediately become positive, leaving very deep cusps at the integrable points. While the values of $\omega$ seem to fluctuate around small positive values for mass ratios other than $1$ and $1/3$, they become progressively more negative exactly at the integrable points as more energy levels are considered, i.e., the minimum continuously moves downward for increasing level numbers used in the statistics as shown in Fig. \ref{massratios}. Here we scan the mass ratio with an even smaller step 0.005. The results indicate that the negative value of $\omega$ is an intrinsic feature of the integrability.

\subsection{Deep learning of wavefunctions}
Through numerical experiments of energy level statistics, we verify theoretical results of previous work and find the statistical difference between integrable and chaotic level spacing distribution, which consumes high computational memory resources. The analysis of level spacing distribution shows a drastic difference between integrable and chaotic systems. We may, on the other hand, extract the information about this integrable-chaotic transition by analyzing the properties of the system wavefunctions. In doing this, we aim to reduce the amount of computation so that the method can be applied to other physical systems. We thus apply the deep learning methods to analyze the wave functions of highly excited states, which contains too complicated information for our eyes to identify. We have to rely on a few known signatures of the integrable states and chaotic states. The chaotic states are expected to resemble a random superposition of plane waves, while the wave functions of integrable states are expected to have some non-trivial morphology.

\subsubsection{Dataset}
\begin{figure}
  \centering
  \includegraphics[width=8.6cm]{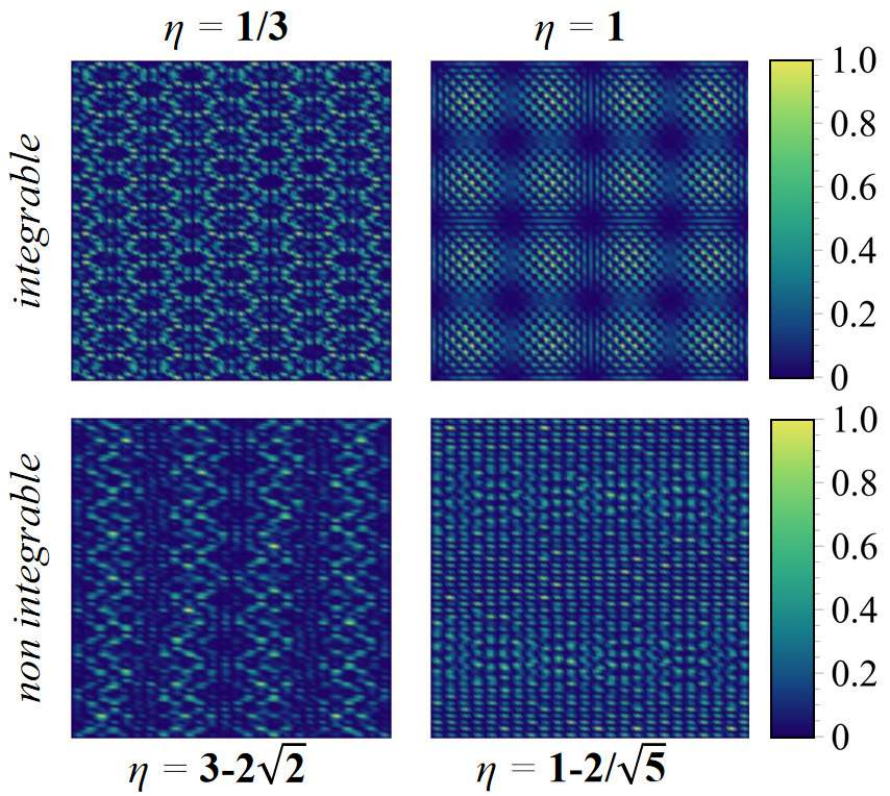}
  \caption{{Probability density diagrams of the $N_L=2000$ state of wave function at different mass ratios. The top two panels show the density plots of integrable systems, while the bottom shows the non-integrable ones. The color from deep to light represents the probability from low to high.}}\label{dataset}
\end{figure}

Deep learning is a data-driven method which requires lots of data. In the field of condensed physics, the methods encoding physical information to neural networks can be divided into several types, including experiments \cite{Gubernatis2018}, Monte Carlo simulation \cite{Juan2017,Wang22017,Huang2017}, and discretization of wave functions \cite{Kharkov2020,Huber2021}, etc. The wave functions of lower excited states are unworthy to analyze \cite{Huber2021}, thus we choose to include the wavefunctions of the $101$th to $2100$th excited states as the input of NN. Both the wave function $\psi(x_1,x_2)$ and the probability density $|\psi(x_1,x_2)|^2$ are good candidates as neural network input, here we choose the latter and present as an example the probability density of the $2000$th state for different mass ratios in Fig. \ref{dataset}. 

To produce dataset, we fix the interaction strength $g=20$ and diagonalize the Hamiltonian (\ref{HamiltonianD}) and plot the probability density $|\psi(x_1,x_2)|^2$ of the $101$th to $2100$th excited states for $\eta=1,1/3$ (with label: integrable) and $\eta=3-2/\sqrt{2},1-2/\sqrt{5}$ (with label: non-integrable). And then the dataset was randomly selected and divided into three parts: training set, test set and validation set, which accounted for $80\%$, $10\%$ and $10\%$ of the total dataset, respectively. We fix the random seed used for drawing to avoid discrepancies between different realizations of the network. The main objective of this work is to verify the feasibility of neural networks in the mass imbalance problem. 

The wave functions $\psi(x_1,x_2)$ are direct product of the eigenvetors of the Hamiltonian with the basis $\psi_{ij}(x_1,x_2)$ in (\ref{basis}). The density plots $|\psi(x_1,x_2)|^2$ are thus continuous functions of the variables $x_1$ and $x_2$, which need to be discretized before being used as a network input. By taking points at intervals in the $x_1$ and $x_2$ directions we can represent the wavefunctions as images with different resolutions. For example, taking spatial intervals of $0.002$ will give a high resolution density map of $500\times500$. This resolution contains too many pixels and the balance between resolution and computational cost deserves to be taken into account. Too high resolution can dramatically reduce computational efficiency, while too low resolution may not capture oscillations in highly excited states, leading to the loss of physical information. The choice of at least a $2\sqrt{N_L}\times2\sqrt{N_L}$ representation of the wavefunction is necessary ($N_L$ denotes energy level number), according to the Nyquist–Shannon sampling theorem \cite{shannon1949,gonzalez2009}. For the excited state $N_L=2100$, at least a $92\times92$ pixel map is needed to represent the wavefunction pattern.

We noticed that there are three methods in the dataset production for network training. The coarse-grained approach focuses more on the local features of the wavefunction \cite{Huber2021}, the interpolated approach \cite{prajapati2012} takes  into account the correlation between neighboring pixels, and direct discretization of the wavefunction at larger spatial scales, without resorting to dimensional reduction of the high-resolution image, optimizes wavefunction integrity and computational cost. For example, a discrete interval of $0.002$ ($500\times500$ pixels) will consume more computing resources (image generation as well as network training) than a discrete interval of $0.01$ ($100\times100$ pixels). Here we generate $100\times100$ pixel probability density maps using $0.01$ intervals and apply them in all experiments below. Our numerical experiments demonstrate that none of the three methods lose the physical information of the wavefunctions at this dataset resolution. This resolution is thus chosen to prevent image distortion, which would destroy the symmetry of the integrable states, while the non-integrable states are are more robust to the noise and some random noise does not change the classification of the network. 

\subsubsection{Neural network}
CNN has been proved to be very capable of extracting features from images through convolution kernel. It avoids complicated preprocess of images and is widely used in fields like object detection and semantic segmentation \cite{Girshick2014,Girshick2015}. After convolution, original images are converted to feature maps and become the input of the pooling layer. After convolution and pooling, all the feature maps are flattened as a 1D vector through a fully connected layer. Finally Softmax function process the 1D vector and output class of every image.

\begin{figure}[ht]
  \centering
  \includegraphics[width=8.6cm]{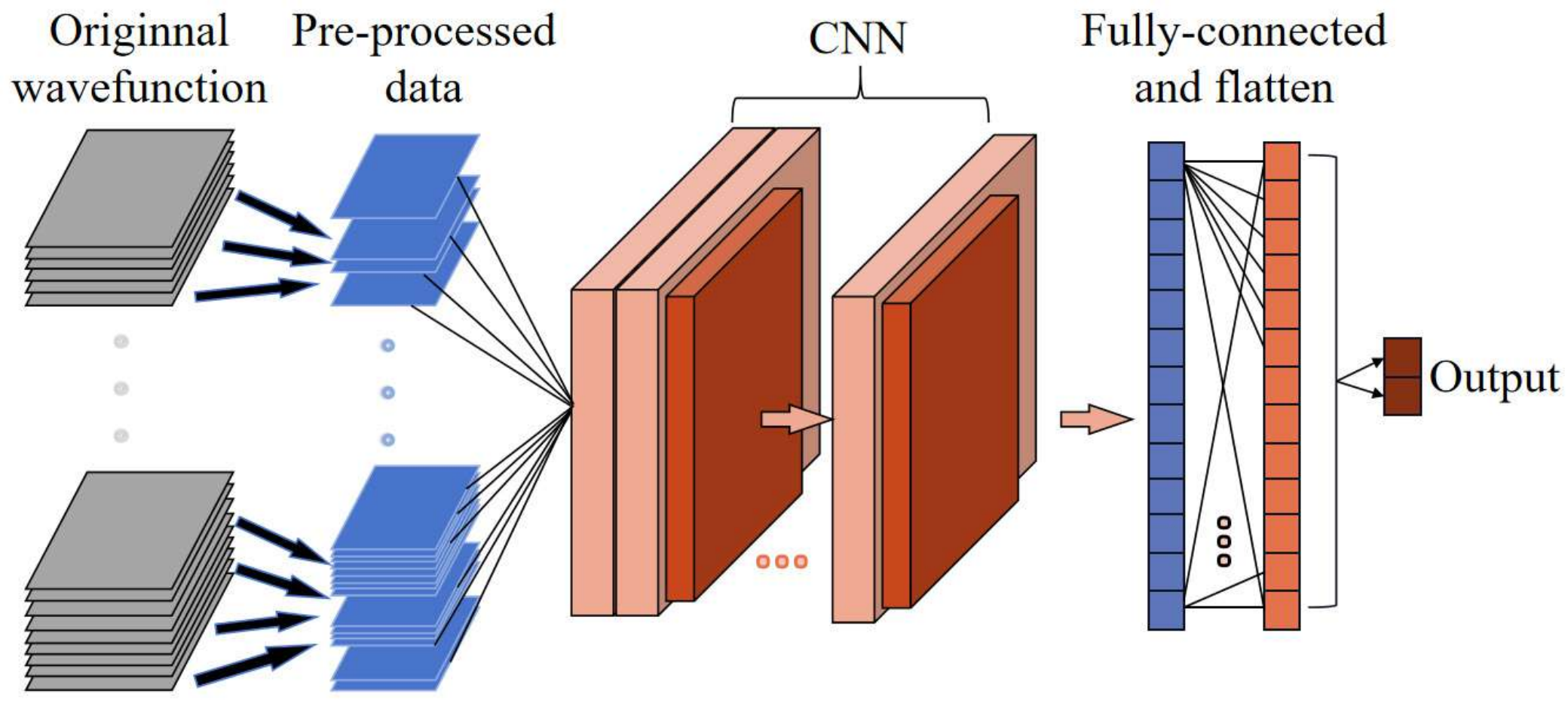}
  \caption{{Schematic of the convolutional neural network used in our analysis. The collection of wavefunction images  is preprocessed (such as labeling, dividing the dataset) and then fed into a classical convolutional neural network, which contains convolutional layers, pooling layers and a fully-connected layer. The last layer produces an output through the activation function ReLu, consisting of two neurons that output the probability of integrable and non-integrable respectively}}\label{CNN}
\end{figure}

We construct a convolutional neural network to distinguish whether the system state is integrable and find the transition point from integrable to chaos, which is intrinsically a binary classification problem, as shown in Fig. \ref{CNN}. The output of the network is a vector with $p$ elements, i.e. $p$ output neurons are typically used to classify $p$ classes. Here we use two neurons corresponding to the integrable and non-integrable state, respectively. More neurons will be applied to the multi-classification case when the method is generalized to discriminate more distinct phases of matter or system states.

The two elements $y_1, y_2$ of the output layer vector represent the probability of the state being integrable and non-integrable respectively, and their sum is $1$. An input state is categorized as integrable if $y_1 > y_2$, and non-integrable if $y_1 < y_2$, a process accomplished by means of either a sigmoid or softmax activation function.

The network aims to study the feature of probability density, which defines a map $y=f(\rho;\theta,\theta_{hyp})$, where $y=\{y_1,y_2\}$ is the output of classification, $\rho=|\psi(x_1,x_2)|^2$ is the images of input probability density, and $\theta_{hyp} $ is the collection of hyperparameters including learning rate, loss function, network structure, etc. The network weight $\theta$ will be adjusted iteratively in the optimization algorithm of the neural network, such as SGD, Adam \cite{Kingma2015}. The $f$ is updated by $\theta$ continuously by training the neural network to get the best fit. The process of adjusting hyperparameters involves the interpretability of neural networks, which is described in \cite{Bau2017}. Deeper neural networks can be used to further improve the accuracy, such as ResNet\cite{He2016}, Inception\cite{Szegedy2016}, etc., which also incurs greater computational overhead. We tested different network structures and obtained no significant difference in the results. Therefore we do not discuss the different network architectures here.

\subsubsection{Numerical experiments}

\begin{figure}[t]
  \centering
  \includegraphics[width=8.6cm]{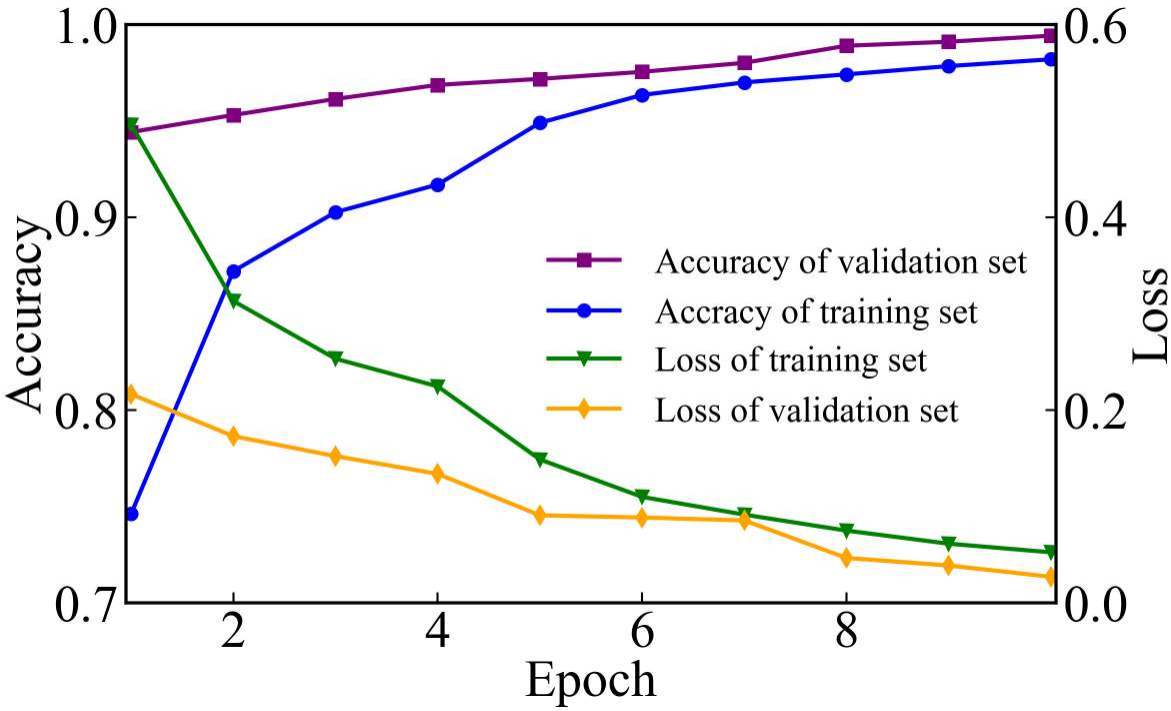}
  \caption{{The accuracy and loss of the neural network training. Different curves represent the accuracy/loss at different stages. The network is trained using numerical simulation data for $2$ integrable and $2$ non-integrable mass ratios with each sample a $100\times100$ pixel image.}}\label{training}
\end{figure}

Following the discussion above, we train and test the CNN and plot the results for the training process in Fig. \ref{training}. Accuracy, the proportion of all samples that a model correctly classifies, is a common metric used to evaluate the performance of classification models. It is usually defined as $Accuracy=N_{correct}/N_{total}$, where $N_{correct}$ is the number of samples correctly classified by the model and $N_{total}$ is the total number of samples performed by the model. Loss is a metric used to measure the difference between the model's predictions and the actual target. It is a measure of the model's performance during training and is often minimized to make the model fit better to the training data. We use binary cross entropy as the loss function, defined as
\begin{equation}
    J(\theta, \rho, y)=-\frac{1}{p} \sum_{i=1}^p\left(y_i \log \left(\hat{y}_i\right)+\left(1-y_i\right) \log \left(1-\hat{y}_i\right)\right),
    \label{Loss}
\end{equation}
where $y_i$ is the actual category of the sample (0 or 1), $\hat{y}_i$ is the probability that the network predicts the sample category to be $y_i$, and the output size $p=2$ in our case. We see from Fig. \ref{training} the accuracy increases with the number of epochs, while the loss averaging all samples $Loss=\sum_{n,\eta} J(\theta, \rho, y)/N_{total}$ deceases, as indicated on the right of Fig. \ref{training}. We used states with mass ratios $\eta=1, 1/3, 3-2\sqrt{3},1-2/\sqrt{5}$ and level numbers from $n=101$ to $2100$ as input data for the network training. After $10$ epochs the network can successfully classify samples from the validation/training set with a high accuracy at 99\%.

\begin{figure}[t]
  \centering
  \includegraphics[width=8.6cm]{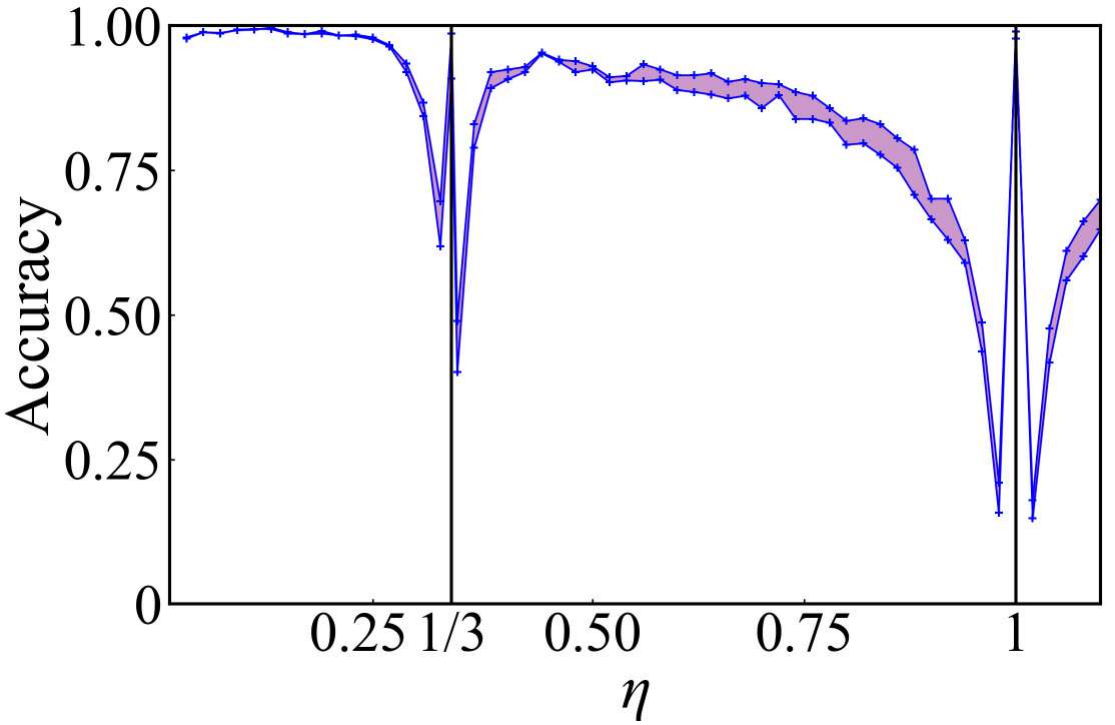}
  \caption{{The upper and lower bounds of the accuracy of the neural network under different random seeds. The purple shaded area represents the fluctuation range of network accuracy, which has a sudden drop near the two integrable points indicated by the black vertical lines.}}\label{predict}
\end{figure}

The randomness in the network leads to different accuracy under different random seeds for the same group of data. The purpose of setting random seed is that the initial parameters in neural network are usually initialized into random numbers, and the local optimal solution of the gradient descent method is very sensitive to the choice of initial position points. Random seed specifies the initial random value, which means the parameters of the network are the same in each operation experiment. Fig. \ref{predict} shows the upper and lower bounds of the accuracy of the neural network under different random seeds, and the amplitude of the fluctuation is represented by the shaded area. The states are classified for other mass ratios by scanning at $0.02$ intervals between $\eta=0$ and $1$. For $\eta<0.25$, we find the accuracy is very close to 1 with quite small fluctuation. Across the full region of mass ratio, there exist two integrable-chaotic transition points, while the accuracy in the region between $\eta=1/3$ and $1$ exhibits strong fluctuations with decreasing accuracy. Note that there is a sudden drop near the two transition points, which indicates that it is difficult to classify the non-integrable states near the integrable mass ratios.

\begin{figure}[t]
  \centering
  \includegraphics[width=8.6cm]{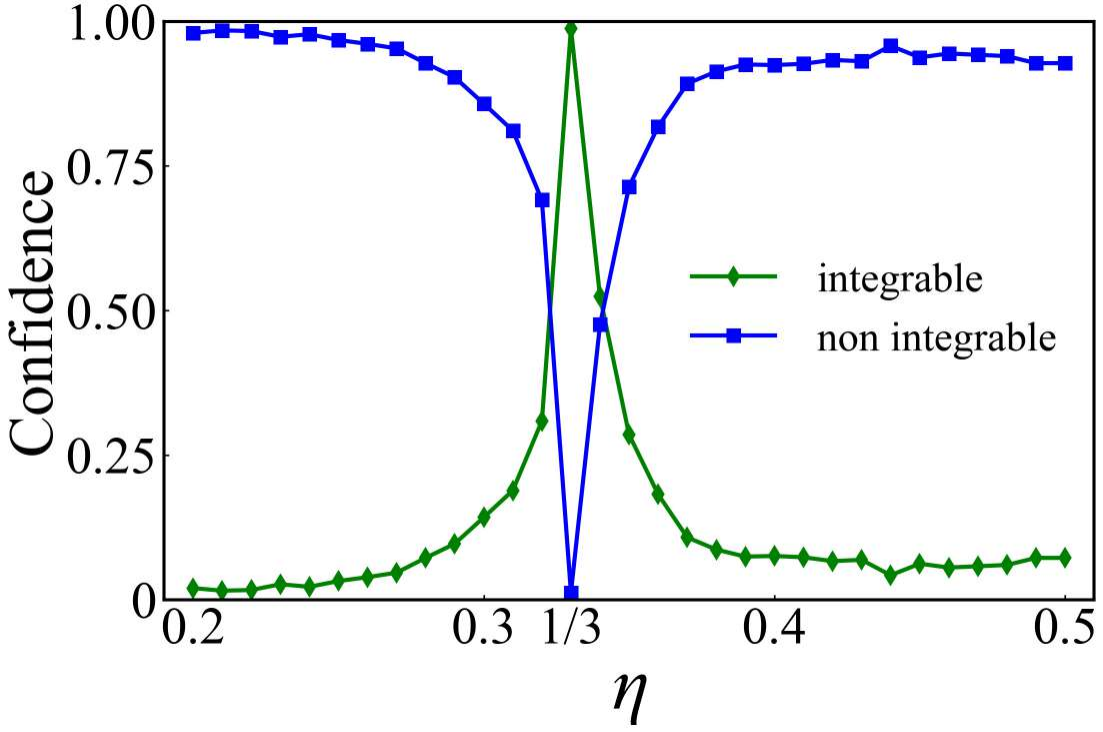}
  \caption{{The blue curve represents the probability that the network considers integrable, while the green curve represents the probability that is non-integrable. At $\eta=1/3$, the probability that the network considers the system integrable reaches 98.78$\%$}.}\label{confidence}
\end{figure}

We expect that our neural network can learn features from the wave function and find the transition between integrable and non-integrable states. To this end, we only include the probability densities for one integrable mass ratio, say $\eta=1$, and one non-integrable one, say $\eta=1-2/\sqrt{5}$, in the training data set. The network trained with these data is used to recognize the transition point between the integrable and chaotic systems. The predict results of NN are shown in Fig. \ref{confidence}. Confidence represents the probability values corresponding to each category output by the network when performing classification. For example, for a cat and dog image classification model, a confidence level of 0.8 for cat samples indicates that the network has $80\%$ probability that the sample is a "cat" and $20\%$ probability that the sample is a "dog". The integrability confidence of our network at $\eta=1/3$ reaches $98.78\%$, which means NN believes that the system is integrable for mass ratio $1/3$. Note that the density probability images of $\eta=1/3$ are not included in the training data. This means that the neural network we built has the ability to extract typical features from the probability density images of some specific mass ratios to recognize the integrability of the system with other mass ratios.

\subsection{Adversarial learning}

Our network has achieved great success in the above experiments. However, current studies show that the accuracy of the neural network will be greatly reduced when a small perturbation is added \cite{Szegedy2013,Goodfellow2015,Xu2020}, which indicates the vulnerability of the network to adversarial attacks. In quantum many-body problems, the perturbations can be manifold. Both traditional machine learning and deep neural networks are vulnerable to perturbations, and it has been shown that both classical and quantum classifiers can be affected by perturbations.  \cite{Lu2020,Jiang2023}.

Adversary samples are those that have perturbations mixed in with the original sample, which may cause classification errors in high-precision networks. The process of generating adversarial samples attack network is called adversarial attack. Adversarial attack can be divided into targeted attack and non-targeted attack in terms of expectation, and can be divided into black-box attack and white-box attack in terms of understanding. Black box attack means that the accuracy of the network is affected when the network parameters or inputs are unknown, while white box attack is carried out when all network parameters are known. Many researches have been made on adversarial learning for improving the security of the network \cite{Biggio2018,Ling2011}. This approach is also applied in the field of quantum many-body physics \cite{Lu2020,Jiang2023,Liao2021}. We expect to improve the neural network to better study wave function morphology and special quantum states, such as quantum scar state \cite{KeskiRahkonen2019}. On the other hand, to increase the robustness of the network to perturbations, we generate adversarial samples to attack our network, and then increase the robustness of the network through adversarial training. We added perturbations by different methods and carried out comparative experiments and analysis.

\subsubsection{Standard perturbation}
We first consider the standard perturbation mixed directly in the probability density which is widely used in image classification. The approaches for adversarial attacks include attacks based on model gradient, optimization and generate adversarial network (GAN). We aim to analyze the morphological characteristics of the probability density learned by the neural network and improve the accuracy and robustness of the network, so we employ the robust defense through adding adversarial samples in the training set, which greatly improves the resistance of the network to perturbation.

Fast gradient sign method (FGSM) is a method of adversarial attacks based on model gradient \cite{Goodfellow2015,Madry2018}. Contrary to the neural network minimization loss function, FGSM aims to maximize the loss by learning network gradient information. To produce adversarial samples $\tilde{\rho}$, it is conventionally to introduce a perturbation in the image dataset of the input probability density ${\rho}$ using the following approach
\begin{equation}
    \tilde{\rho}=\rho+\varepsilon \operatorname{sign}\left(\nabla_{\rho} J(\theta, \rho, y)\right),
\end{equation}
where $\nabla_{\rho}$ is gradient of the network over the matrix $\rho$, which is solved by back-propagation method \cite{Y1998}. The sign function changes the value greater than $0$ to $1$ and the value less than $0$ to $-1$. $\varepsilon$ is a small coefficient that controls the strength of the perturbation. 

\begin{figure}
    \centering
    \includegraphics[width=8.6cm]{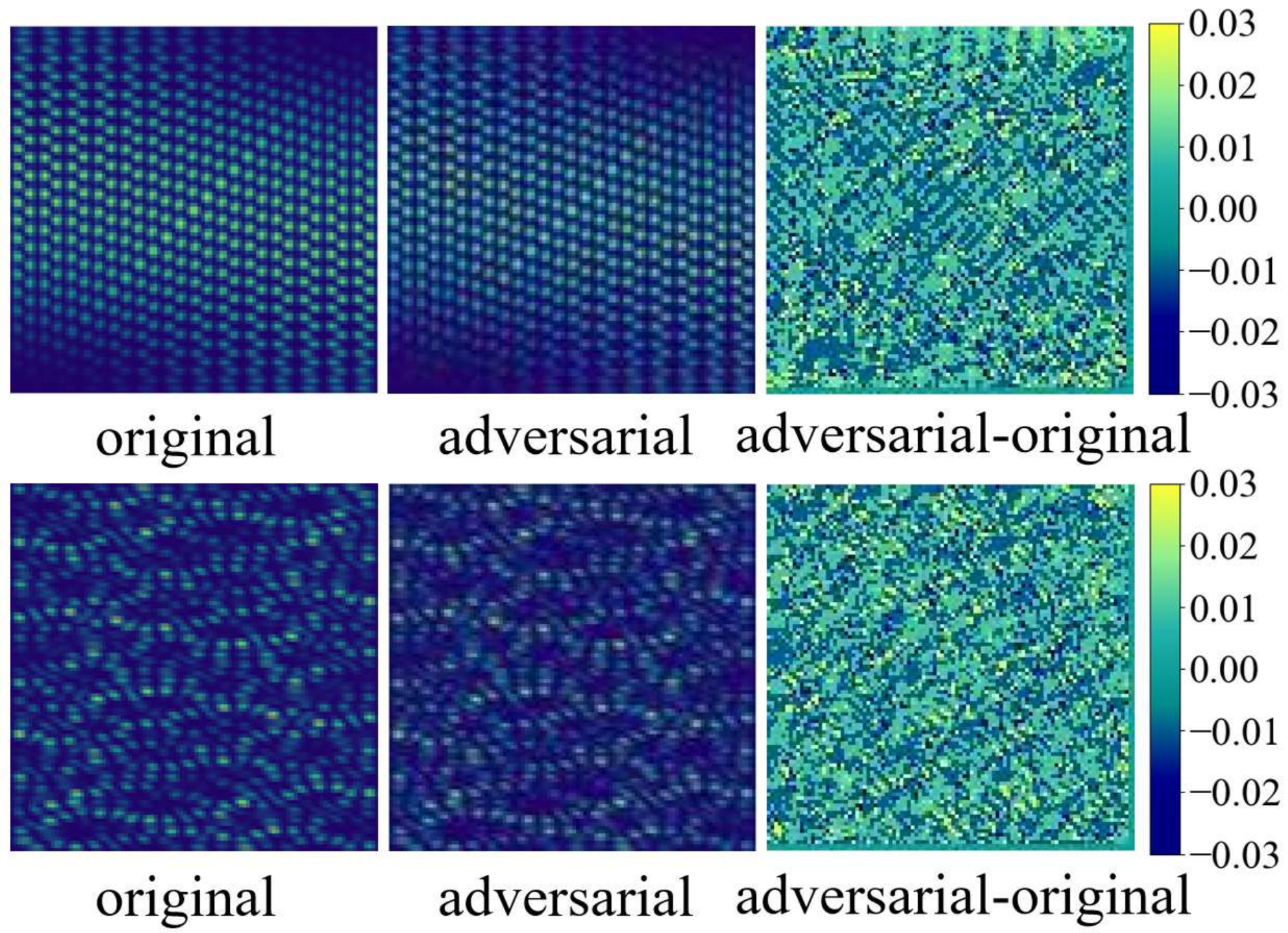}
    \caption{{Schematic diagram of adversarial samples. The first column shows the original images $|\psi|^2$ for the $960$th state with mass ratios $\eta=1/3$ (integrable) and $\eta=3-2\sqrt{2}$ (non-integrable), and the network correctly recognizes them as integrable and non-integrable, respectively. The second column shows the adversarial samples generated by the addition of perturbations to $|\psi|^2$, which leads to misclassification by the network. The third column shows the pixel difference between the adversarial samples and the original ones. The colorbar corresponds only to the images in the third column in the case of $\varepsilon=0.03$.}}\label{adversarialsample}
\end{figure}

Using FGSM, we add perturbations to the probability density images. The perturbations we add cause the predictions of the network to change completely, while the human eye barely detects any change. This change is easy to realize for both integrable and non-integrable states. Schematic examples of adversarial examples are given for integrable and non-integrable mass ratios in Fig. \ref{adversarialsample}. The middle column shows the adversarial images generated by the addition of perturbations into the original probability densities. A slight change in each pixel value would confuse the network and result in a false classification.

\begin{figure}
    \centering
    \includegraphics[width=8.6cm]{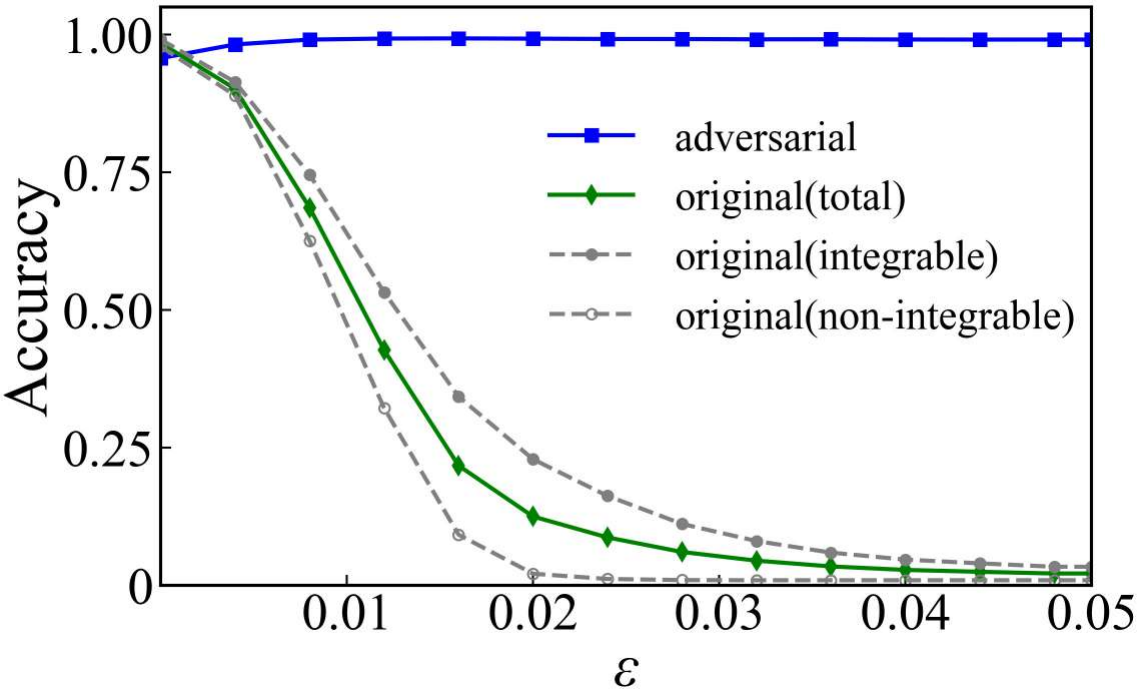}
    \caption{{The results of robust adversarial learning under standard perturbation. The blue line represents the accuracy of the network after adversarial learning, while the green line represents that of the original network. The accuracy for integrable states (solid dots) and non-integrable states (open circles) are shown in gray. By adding adversarial samples to the training set, the resistance of the neural network to perturbation is greatly enhanced after training.}}\label{adversarial}
\end{figure}

Perturbations may have dramatic effect on the morphology of the wave function. Too high a perturbation makes the network more confused but changes the original morphology of the wave function, while too low a perturbation makes it difficult to disturb the network. We limit the perturbation to an interval $(-\varepsilon, +\varepsilon)$ that ensures the adversarial samples successfully interfere with the network without damaging the wave function morphology. Adversarial training is considered as one of the effective methods to enhance model robustness \cite{Madry2017}. However, after adversarial training, the performance of neural networks tends to degrade when dealing with clean samples, thus the trade-off between model accuracy and robustness needs to be carefully considered \cite{Wen2020}. After generating adversarial samples, we use adversarial training to improve the network robustness, as shown in Fig. \ref{adversarial}. The robustness of the model to perturbation is greatly improved for a perturbation strength up to $\varepsilon \sim 0.05$.

There is no universally effective adversarial defense method that can protect against all possible adversarial sample attacks. In fact, different strategies need to be employed to enhance the network's resilience against different types of attacks. While certain adversarial defense methods may demonstrate robustness against specific types of adversarial attacks, it remains challenging to address all possible attack scenarios. Therefore, a combination of multiple defense strategies is necessary, including adversarial training, model fusion \cite{tramer2017}, randomness enhancement \cite{luo2016}, etc. In quantum theory, the perturbation correction arising from the Hamiltonian term is applied to the wavefunctions rather than to the probability densities. In this sense, the vulnerability of neural network to recognize chaotic states through an adversarial learning approach should be protected against the quantum perturbations on the wave function, which will be done in the following subsection.

\subsubsection{Quantum perturbation}
\begin{figure}
    \centering
    \includegraphics[width=8.6cm]{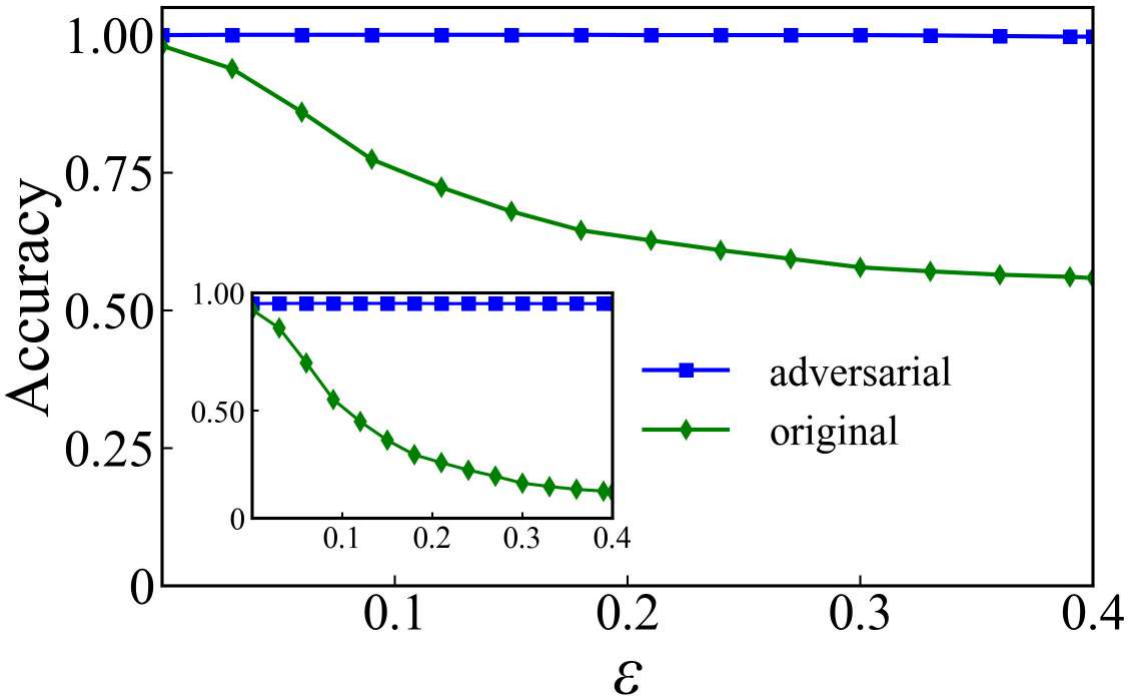}
    \caption{{The effect of quantum perturbations on the network. The blue line represents the accuracy of the network after adversarial learning, while the green line represents that of the original network. Inset: the accuracy for the integrable states only, while the accuracy for non-integrable states remains largely unaffected. Following adversarial learning, the network demonstrates significant improvement in accuracy for both all states and integrable states.}}\label{result_qad}
\end{figure}

Now consider that the perturbations are applied on the wavefunctions and the probability densities as input dataset are distorted accordingly. We generate quantum adversarial samples by perturbing the integrable wavefunctions by non-integrable data for the adversarial training, and vise versa. For all excited states, we obtain the adversarial samples $\tilde{\rho}$ by mixing small perturbation into the original wavefunction as follows
\begin{equation}
    \tilde{\rho}=|\psi(x_1,x_2)+\varepsilon \psi'(x_1,x_2)|^2
,\end{equation}
where $\psi'(x_1,x_2)$ is the wave function which belongs to the category other than $\psi(x_1,x_2)$. Predictions of the network for the states from the dataset with wavefunction perturbation is shown in Fig. \ref{result_qad}. Similar to the standard perturbation scheme in the probability density, the accuracy of the network in quantum perturbation can be greatly enhanced by  adversarial training. Small values of $\varepsilon$ lead to weak noise and the original network is capable of correctly classifying all input states, while the network fails for larger values of $\varepsilon$ which leads to confusing input states. It actually fails for integrable states where the noise destroys the correlation and symmetry of the states, as shown in the inset of Fig. \ref{result_qad}, and the accuracy for non-integrable states remains almost constant with the increase of disturbance. Our network tends to categorize states without clear spatial correlations but with nodal lines as non-integrable. In essence, the network predominantly identifies most states as non-integrable, with only a few instances recognized as integrable states. This asymmetry in learning is not typically observed for labels of equal significance, such as in the cat and dog example. Integrable states exhibit a greater abundance of features, and their symmetry may be disrupted during the perturbation process. On the other hand, non-integrable states display irregular particle motion in phase space, making them less susceptible to quantum perturbations.

It is necessary to compare the two adversarial learning results on the network as we generate samples by standard and quantum perturbations respectively. Standard perturbations completely change the predictions of the network for very small $\varepsilon\sim 10^{-2}$, which happens equally for both the integrable and non-integrable states. For quantum perturbations, on the other hand, the network is still able to correctly categorize the majority of states for perturbations generated by $\varepsilon$, e.g. the accuracy stays over $50\%$ for $\varepsilon \sim 0.4$, which is an order of magnitude stronger compared with the standard perturbation. This suggests that the network is more vulnerable to perturbations added to the probability density, while it exhibits greater resilience or robustness to perturbations added to the wavefunction, particularly in the case of non-integrable states. As the input states become chaotic, the network's accuracy in classifying integrable states gradually decreases. However, the accuracy for non-integrable states remains consistently high. These numerical experiments generate lots of data, allowing the neural network to be used to recognize atypical states that do not fit the overall pattern, such as scar states \cite{Stockmann1990}. With adversarial learning, the network demonstrates significant improvement in accuracy for both all states and integrable states, even in the presence of strong perturbations.

We will not go into the details on the impact of different adversarial attack algorithms on the network accuracy. Interesting future works include, for example, the generative adversarial network (GAN), which generates intrinsic adversarial samples through generator mapping. Taking the wave function itself as input instead of the image can help the neural network learn more intrinsic new physics in quantum many-body systems. The adversarial learning may be applied to the field of quantum physics and find a better defense method. The robustness of neural networks and interesting phenomena in many-body physics can be achieved by adversarial sampling in the wavefunctions.

\section{Conclusion}
In conclusion, our study focuses on the application of deep learning neural network in 1D many-body physics of two mass-imbalanced atoms in a hard-wall trap. Through careful examination of the energy level statistics, we highlight the differences in the energy spectrum and energy level distributions for integrable and non-integrable systems, respectively. Especially, the levels for integrable models are exactly constants. i.e., independent of interaction, which serve as perfect benchmarks for numerical accuracy estimation. The numerical data of density distribution produced from ED method is still within the order of $10^{-3}$ for high level number around $5000$ and large interaction strength $g = 50$. The level spacing distributions are fitted by a Brody distribution with the fitting parameter $\omega$ and we observe the existence of a critical line that separates integrable and non-integrable points, which is determined by $\omega=0$. Our findings confirm the Bethe-ansatz integrable mass ratios $\eta=1, 1/3$ and do not reveal any new integrable points.

We have built a convolutional neural network of $100\times 100$ pixels probability density images with output vector with 2 elements. Based on the morphological analysis of probability density, we find that deep learning of wavefunctions proves to be a very effective way to determine the transition between integrable and chaotic states when data is sufficient. We have successfully identified the transition points between integrable and non-integrable systems with a much shorter computation time compared with the energy level statistics method, which reaches high accuracy in a few iterations. A series of numerical experiments indicated that deep learning can effectively learn the morphology of wave functions to discern the integrability of a physical system, even without prior knowledge of its underlying physics. An exemplary instance is the network's ability to identify a new integrable point $(\eta=1/3)$ by leveraging knowledge from other known integrable mass ratios $(\eta=1)$, achieving a remarkable network confidence of $98.78\%$. This showcases the network's capacity to generalize and make accurate predictions beyond the explicitly provided data. 

We further improved the robustness of our neural networks through adversarial learning by generating the samples by two perturbation schemes. For very small perturbation $\varepsilon\sim 10^{-2}$, standard perturbations have a profound impact on the network's predictions, leading to a complete alteration of its outputs. This effect is observed equally for both the integrable and non-integrable states. Under quantum perturbations the accuracy of the original network remains consistently above $50\%$ for strong perturbations around $\varepsilon \sim 0.4$. Interestingly, this phenomenon primarily affects the accuracy of the network's predictions for integrable states, while the accuracy for non-integrable states remains relatively unaffected. Either way, the network accuracy has been significantly improved by adding
adversarial samples to the training set. Adversarial learning holds great promise in the field of many-body physics, particularly in the study of other integrable models and quantum chaos \cite{Victor1987,McDonald1988,Alves2020}.

\begin{acknowledgments}
This work is supported by National Natural Science Foundation of China (NSFC) under Grants No. 12074340, No. 12105245 and No. 12204406, and the Fundamental Research Funds of Zhejiang Sci-Tech University (Grant No. 21062106-Y). 
\end{acknowledgments}

\nocite{*}

\bibliography{reference}

\end{document}